\documentclass[a4paper]{jpconf}
\usepackage{graphicx}
\begin{document}
\title{Spin-spin Relaxation Time Measurements of 2D $^3$He on Graphite}

\author{D Sato, K Naruse, T Matsui and Hiroshi Fukuyama}

\address{Department of Physics, Graduate School of Science, The University of Tokyo,
7-3-1 Hongo, Bunkyo-ku, Tokyo 113-0033, Japan}

\ead{sato@kelvin.phys.s.u-tokyo.ac.jp, hiroshi@phys.s.u-tokyo.ac.jp}

\begin{abstract}
Spin-spin relaxation time ($T_2$) and magnetic susceptibility ($\chi$) of the 
second layer $^3$He adsorbed on Grafoil, exfoliated graphite, 
preplated with a monolayer $^4$He 
are studied by pulsed-NMR in a density range of $0.68 \leq \rho \leq 5.28$ 
nm$^{-2}$. The temperature dependence of $\chi(T)$ and $\chi(T = 0)$
show Fermi fluid behaviour and no evidence of self-condensation 
are found even at the lowest density $\rho = 0.68$ nm$^{-2}$.
Density dependence of $T_2$ at $f = 5.5$ MHz shows a broad maximum of 
5.7 ms around $\rho = 3$ nm$^{-2}$. 
Since the decrease of $T_2$ in dilute side can not be expected in the ideal 2D fluid,
it can be understood as the relaxation caused by a small amount of
solid $^3$He at heterogeneity of the substrate.
We also measured the Larmor frequency dependence of $T_2$ at $\rho = 5.28$ nm$^{-2}$. 
$1/T_2$ has a $f$-linear dependence similarly to the earlier
study on a first layer solid $^3$He \cite{Cowan87}. From a comparison between our result 
and the earlier one, this linearity is almost independent of the 
particle motion. Now, it could be caused by
a microscopic magnetic field inhomogeneity arisen from the mosaic angle 
spread and diamagnetism of the graphite substrate.

\end{abstract}

\section{Introduction}
The second layer $^3$He on graphite is an ideal example for
studying two-dimensional (2D) Fermion system.
By controlling the number of adsorption atoms, we can widely change
the areal density $\rho$ from nearly ideal Fermi gas to 
highly compressed 2D solid \cite{Greywall90, Fukuyama08}.
At the commensurate phase in the second layer, so called ``4/7 phase",
a gap-less quantum spin liquid state is suggested. 
The transition between a Fermi fluid and the 4/7 phase 
is Mott-Hubbard type
with divergence of the effective mass $m^*$ \cite{Casey03}.
Moreover, recent heat capacity study shows the existence of
a low-density liquid state ($\rho_{liq} \leq 1$ nm$^{-2}$) 
in the third layer \cite{Sato10}. But, this self-condensed 
state has not been confirmed in the second layer yet.
To study the spin correlation in the quantum spin liquid, 
the self-condensed liquid,
and the relaxation process in the liquid or dilute 
Fermi gas, pulsed-NMR is a useful probe.

In this work, we report a preliminary pulsed-NMR study of the 
spin-spin relaxation time ($T_2$) and magnetic susceptibility 
($\chi$) of Fermi fluid phase in the second layer.
To remove the large Curie susceptibility from the first layer,
we replaced the first layer $^3$He with $^4$He. 
It is also helpful to reduce
the substrate heterogeneity effect because the heterogeneous regions
with deep potential are preferentially filled with $^4$He atoms.

\section{Experimental}
The substrate used in this work was Grafoil, an exfoliated graphite,
with the total surface area of $A_{tot} = 53.6$ m$^2$. Grafoil
consists of 10-20 nm size micro-crystallites with a mosaic angle
spread of $\pm 30^{\circ}$ \cite{Takayoshi10}.
Before adding $^3$He sample at 2.7 K, 11.78 nm$^{-2}$ of $^4$He was adsorbed 
at 4.2 K as the first layer. This density is slightly higher than 
the highest density of the first layer $^3$He determined from 
the neutron scattering (11.06 nm$^{-2}$) \cite{Lauter90} and 
the heat capacity measurement (11.4 nm$^{-2}$) \cite{Greywall90}.
Therefore, we believe that $^3$He atoms don't sink into the first layer.
The conventional spin-echo technique with the pulse sequence of
$90^{\circ}$-$\tau$-$180^{\circ}$ were used for measuring the spin-spin relaxation time
$T_2$. In presence of the mosaic angle spread of Grafoil, 
the decay of the spin-echo signal shows about 5\% of
long tail component due to the angular dependence of $T_2$ \cite{Takayoshi09}. 
But, in this work, spin-echo measurements were carried out within $0 \leq t \leq 2T_2$
where the contribution of the long tail is not important,
and $T_2$ was obtained by single exponential fitting.
Most of the experiments, except a Larmor frequency dependence of $T_2$,
were carried out in a Larmor frequency of 5.5 MHz ($B = 172$ mT).
The static magnetic field is parallel to the substrate. 
The details of this set-up are described in earlier paper, 
where Grafoil is replaced with ZYX \cite{Murakawa03}.

\begin{figure}[b]
\begin{center}
\includegraphics[width=35pc]{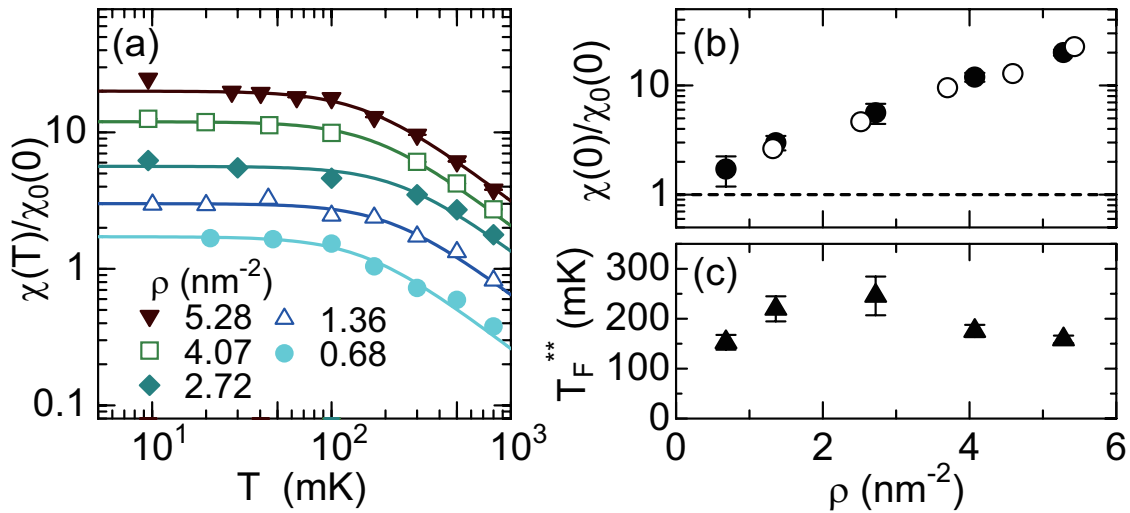}\hspace{1pc}
\caption{\label{fig1}
{\bf a.} Temperature dependence of $\chi(T)/\chi_0(0)$. The solid curves are
fitting lines by eq.(\ref{eqn1}).
{\bf b.} Density dependence of the susceptibility at $T = 0$.
The closed circles are the data taken in this work, while the open
circles are those in ref.\cite{Lauter90}.
{\bf c.} Density dependence of effective temperature $T_F^{**}$.}
\end{center}
\end{figure}

\section{Results and discussions}
Figure 1(a) shows the temperature dependence of the magnetic susceptibility 
$\chi(T)$ at $^3$He density range of $0.68 \leq \rho \leq 5.28$ nm$^{-2}$.
The signals are well described by the phenomenological expression for
the susceptibility of a Fermi liquid \cite{Dyugaev90}:
\begin{equation}\label{eqn1}
\chi(T) = C/\sqrt{T^2 + T_F^{**2}}
\end{equation}
down to $\rho = 0.68$ nm$^{-2}$, which is about half of 
the lowest density in the earlier study \cite{Lusher91}.
Now, $C$ is the Curie constant and $T_F^{**}$ is the effective Fermi
temperature.
At $\rho = 5.28$ nm$^{-2}$, $\chi$ increases slightly at lowest temperature $T \approx 10$ mK. 
Considering a small amount of paramagnetic solid $^3$He, this fraction corresponds to 
within 2\% of the total $^3$He atoms in experimental accuracy.
We thought that this is trapped $^3$He atoms in substrate heterogeneities, and it can be
removed by introducing a few percent of excess $^4$He atoms \cite{Lusher91, Collin01}.

In Landau Fermi liquid theory, the $T = 0$ susceptibility of 2D Fermi liquid 
is given by $\chi(0) = C/T_F^{**} = (m^*/m)(1+F_0^a)^{-1}C/T_F$, where 
$F_0^a$ is a Landau parameter and $T_F$ is the Fermi temperature. 
Since $C$ is proportional to $N$, number of $^3$He atoms, and
$T_F$ is proportional to $N/A$, where $A$ is the surface area
over which the $^3$He liquid occupies,
we obtain $\chi(0) \propto A(m^*/m)(1+F_0^a)^{-1}$.
Considering the self-condensation, 
$A$ would be smaller than the total surface area ($A \leq A_{tot}$).
Now, because the $T = 0$ susceptibility of an ideal 2D Fermi gas is 
$\chi_0(0) = C/T_F \propto A_{tot}$, 
we obtain $\chi(0)/\chi_0(0) = (A/A_{tot})(m^*/m)(1+F_0^a)^{-1}$.
If $^3$He atoms are uniformly spread over the surface area ($A = A_{tot}$),
this function approaches to a constant value (in ideal gas case, unity) at dilute limit. 
On the other hand, when self-condensation occurs, $A$ changes linearly with increasing $N$,
resulting in $\chi(0)/\chi_0(0) \propto N$ similarly to 
the density dependence of $\gamma$ 
in heat capacity measurement of the third layer \cite{Sato10}.
Figure 1(b) shows obtained $\chi(0)/\chi_0(0)$ against $\rho$.
The value approaches to unity with decreasing $\rho$, 
and there is no anomaly down to $\rho = 0.68$ nm$^{-2}$.
Therefore, there is no evidence of self-condensation 
at least at the density above 0.68 nm$^{-2}$ in the second layer.

The density dependence of $T_2$ at $T = 100$ mK and $f = 5.5$ MHz is
shown in Figure 2. 
It shows a broad maximum of 5.7 ms around $\rho = 3$ nm$^{-2}$.
At high density side, the particle motion is suppressed by
the divergence of $m^*$ toward a Mott localization \cite{Casey03}.
Therefore, $T_2$ decreases with increasing $\rho$ due to
the suppression of motional narrowing.
At the lowest density $\rho = 0.68$ nm$^{-2}$, 
the scattering radius $d$ is much shorter than
a mean free path $\ell \approx 1/2d\rho$,
and a dilute 2D $^3$He gas model is applicable.
From an analogous approach to a dilute bulk $^3$He gas \cite{Chapman74},
we obtain:
\begin{equation}\label{eqn2}
1/T_2 \sim vd\rho(\mu^2/d^6)(d/v)^2
\end{equation}
where $\mu$ is magnetic moment of $^3$He, and the averaged velocity
$v \approx 2v_F/3 \propto \rho^{1/2}$ at Fermi degenerated gas
($T < T_F^{**}$ from Figure 1(c)). 
As a consequence, $T_2$ increases with decreasing $\rho$
by $T_2 \propto \rho^{-1/2}$ at dilute limit.
On the other hand, obtained $T_2$ decreases at $\rho = 0.68$ nm$^{-2}$.
It indicates that there is another relaxation process that becomes important 
at dilute limit. 

\begin{figure}[h]
\begin{minipage}{18pc}
\includegraphics[width=17pc]{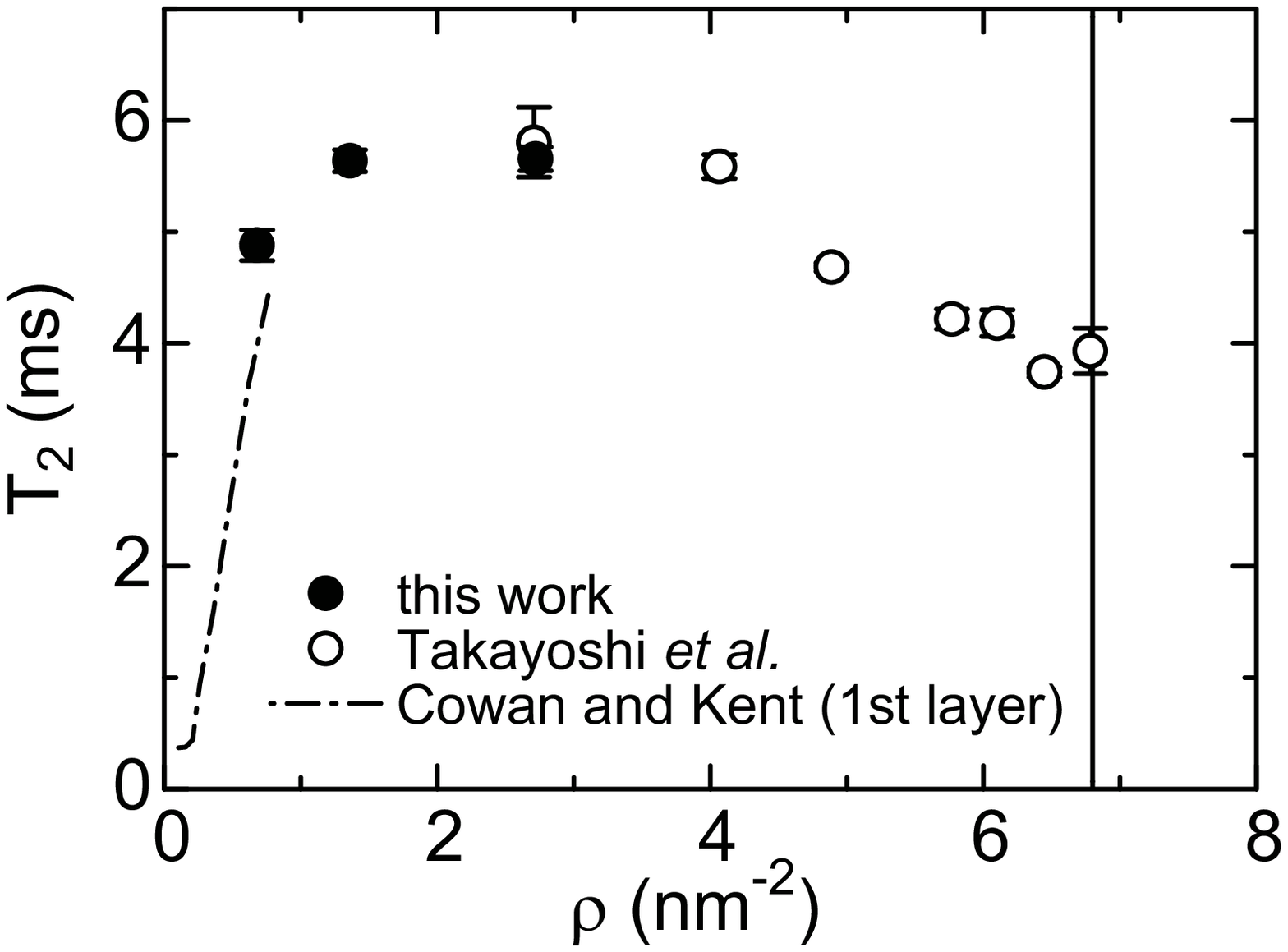}
\caption{\label{fig2}
The density dependence of $T_2$ of the second layer $^3$He at 100 mK and 5.5 MHz.
The closed circles are the data in this work. The open circles are those
in the previous study \cite{Takayoshi09}.
The vertical line corresponds to the density of the commensurate solid
(4/7 phase). 
The dashed-dotted line is a reference in the first layer at 
1.2 K and 5 MHz \cite{Cowan84}.}
\end{minipage}\hspace{2pc}%
\begin{minipage}{18pc}
\includegraphics[width=18pc]{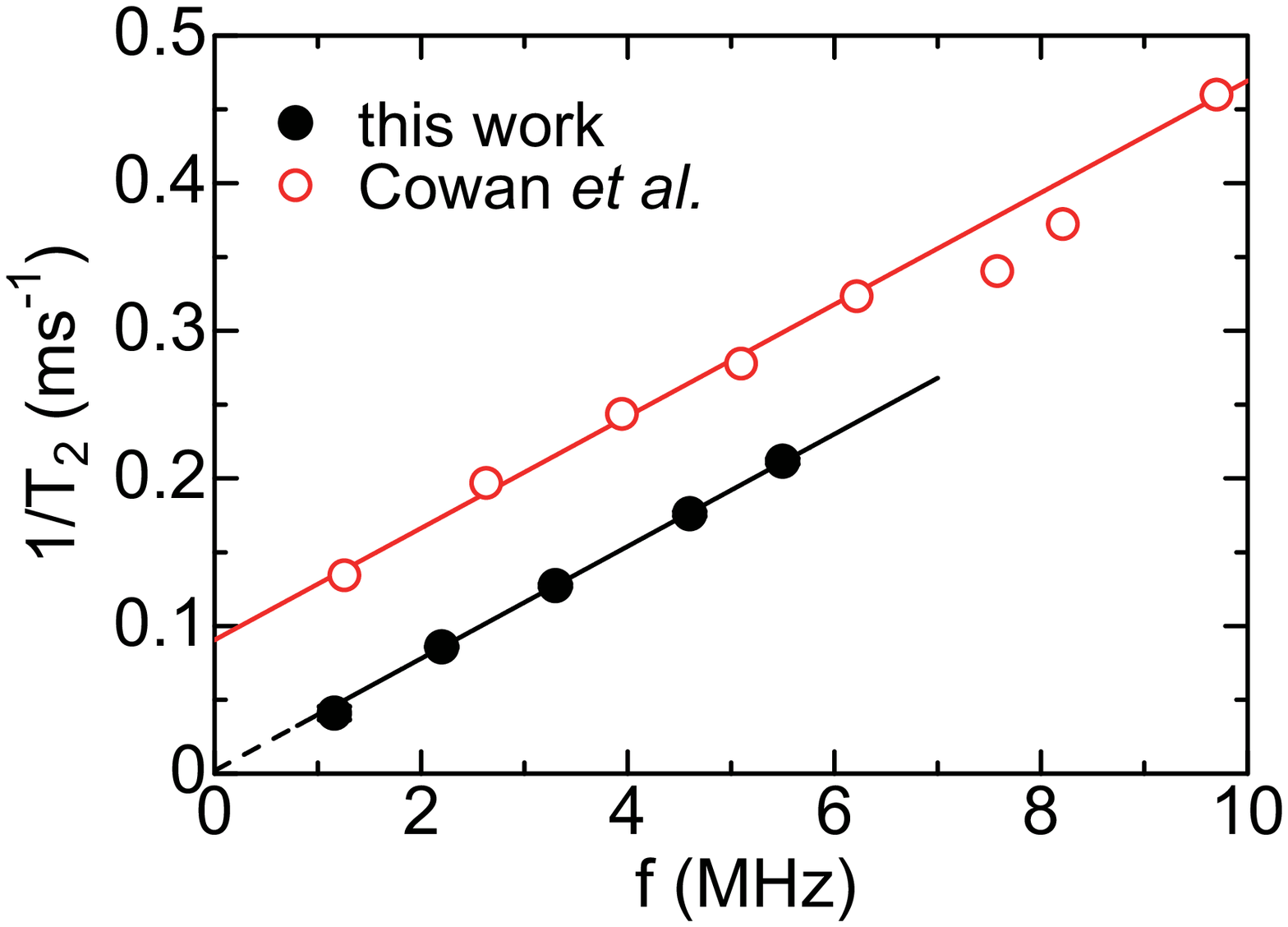}
\caption{\label{fig3}
Larmor frequency dependence of $T_2$. 
The closed circles are $T_2$ of the second layer Fermi fluid at 
$\rho = 5.28$ nm$^{-2}$ and $T = 100$ mK. The open circles are those of
the first layer solid at a coverage of 0.76 monolayer and $T = 1.2$ K
\cite{Cowan87}. The solid lines are linear fitting to the data.}
\end{minipage} 
\end{figure}

Cowan and Kent measured $T_2$ of very low density $^3$He 
films on bare graphite \cite{Cowan84}, and showed that a small amount of
$^3$He ($\sim$ few percent of monolayer) would form high-density solid 
with $T_2 = 0.3$ ms in heterogeneous region of the substrate.
By atomic exchanges between this solid and the 2D gas,
relaxation in the heterogeneous region dominates the relaxation 
process of the 2D gas.
The linear $\rho$-dependence of $T_2$ in the ``first layer'' 
is also shown in the same graph (Fig.\ref{fig2}) for the reference. 
Because the adsorption potential in the second layer is about 
an order of magnitude smaller than that of the first layer,
it is unrealistic to consider such a high-density solid.
However, since the heterogeneous region can exist as a relaxation spot,
and very small amount of $^3$He could also exist as a solid.
This is consistent with the increase of $\chi(T)$ at lowest temperature.
Magnetic impurity in substrate may be another causation of the relaxation.

In Figure 3, we show the Larmor frequency dependence of $T_2$ in a high-density
Fermi fluid ($\rho = 5.28$ nm$^{-2}$) at $T = 100$ mK (closed circles). 
We observed a linear relation between $1/T_2$ and $f$.
This characteristic linear dependence of $T_2$ is also observed in a first layer
solid $^3$He \cite{Cowan87} (open circles).
It is curious since $T_2$ should be independent of $f$ as long as
$f \ll 1/\tau$, where $\tau$ ($\sim d/v \sim 10^{-11}$ sec) is the correlation time.
Although the explanation for this linearity is not satisfactory established, 
this could be a microscopic magnetic field inhomogeneity effect
caused by the mosaic angle spread and diamagnetism of the graphite substrate.
Assuming that the frequency dependence is an extrinsic effect related
to the diffusion process in the microscopic magnetic field gradient,
the intrinsic $T_2$ is obtained by extrapolating to $f = 0$
where the diamagnetic field of graphite disappears.
Namely, $1/T_2 = 1/T_2^{int} + c f$, where $T_2^{int}$ is the 
intrinsic $T_2$ of the system.
From the extrapolation, 
$T_2^{int}$ at $\rho = 5.28$ nm$^{-2}$ is very long ($T_2^{int} > 0.1$ s)
although the strict value is not available because $y$-intercept is too small.
This is much longer than that of the first layer solid ($T_2^{int} \approx 10$ ms).
Nevertheless, the value of $c$, which is a slope of $1/T_2$, 
is almost the same between these two samples.
Therefore, it is clear that the $f$-dependence of $T_2$ is extrinsic one.
To obtain the strict value of $T_2^{int}$, lower frequency experiment or 
a substrate with larger platelet size and smaller mosaic angle will be
required.

In conclusion, we measured $\chi(T)$ and 
$T_2$ of the second layer $^3$He at the density region of 
$0.68 \leq \rho \leq 5.28$ nm$^{-2}$ on graphite pleplated with a
monolayer $^4$He.
Obtained $\chi(T)$ shows Fermi fluid behaviour at all densities, and 
there is no self-condensed state down to $\rho = 0.68$ nm$^{-2}$.
Density dependence of $T_2$ at $f = 5.5$ MHz shows a broad maximum of
5.7 msec. The decrease of $T_2$ at $\rho = 0.68$ nm$^{-2}$ could be
related to the $^3$He solid at the heterogeneity in the substrate.
We also observed a $f$-linear dependence of $1/T_2$, which is similar to
that in the earlier study for the first layer solid \cite{Cowan87}.
This $f$-dependence of $T_2$ is an extrinsic effect, and 
the intrinsic $T_2$ in the Fermi fluid is much longer than measured value.

\smallskip

\end{document}